\documentclass[prl,aps,twocolumn,showpacs]{revtex4}
\usepackage{color,graphicx,bm,amsmath}

\begin{document}

\newcommand{\half}{\ensuremath{\frac{1}{2}}}
\newcommand{\degree}{\ensuremath{^{\circ}}}
\newcommand{\adag}{\ensuremath{a^{\dagger}}}
\newcommand{\bra}[1]{\ensuremath{\langle #1 |}}
\newcommand{\ket}[1]{\ensuremath{ | #1 \rangle}}
\newcommand{\braket}[2]{\ensuremath{\langle #1 | #2 \rangle}}
\newcommand{\trace}[1]{\ensuremath{{\rm Tr}\{#1\}}}
\newcommand{\nbar}{\ensuremath{\overline{n}}}
\newcommand{\nav}{\ensuremath{\langle n \rangle}}
\newcommand{\gbar}{\ensuremath{\overline{\gamma}}}

\newcommand{\gf}[1]{\ensuremath{g^{(1)}(#1)}}
\newcommand{\gs}[1]{\ensuremath{g^{(2)}(#1)}}

\title{Coherence Properties of the Microcavity Polariton Condensate }

\author{D. M. Whittaker}
\affiliation{Department of Physics and Astronomy, University of
Sheffield, Sheffield S3 7RH, UK.}

\author{P. R. Eastham} \affiliation{Department of Physics, Imperial
College London, London SW7 2AZ, UK.}

\date{\today}

\begin{abstract} 
A theoretical model is presented which explains the dominant
decoherence process in a microcavity polariton condensate. The
mechanism which is invoked is the effect of self-phase modulation,
whereby interactions transform polariton number fluctuations into
random energy variations. The model shows that the phase coherence
decay, \gf{\tau}, has a Kubo form, which can be Gaussian or
exponential, depending on whether the number fluctuations are slow or
fast. This fluctuation rate also determines the decay time of the
intensity correlation function, \gs{\tau}, so it can be directly
determined experimentally. The model explains recent experimental
measurements of a relatively fast Gaussian decay for \gf{\tau}, but
also predicts a regime, further above threshold, where the decay is
much slower.
\end{abstract}


\maketitle

Microcavity polaritons are quasi-particles arising from the strong
coupling between excitons and photons confined in planar cavity
structures. The observation of coherent emission from a CdTe
microcavity\ \cite{cond} has demonstrated that polaritons can form a
new type of quantum condensate. As in other quantum condensates, such
as atomic gases or superconductors, a key property is the existence of
an order parameter, the local phase, which is correlated over large
times and distances. The polariton condensate presents interesting
theoretical challenges since, unlike these other systems, it is
mesoscopic, typically consisting of a few hundred particles, and out
of equilibrium, with pumping required to maintain the population
against emission losses. In mesoscopic systems, order parameters
fluctuate\ \cite{callenflucs,eastham:085306}, so the phase
correlations decay. In this letter, we present a theory which shows
that the source of the fluctuations is variations in the number of
condensed particles, induced by the pumping process: it is this
dynamics that is responsible for the decoherence. Our theory shows
that, under appropriate pumping conditions, existing microcavity
structures should display much longer coherence times than currently
measured, opening up opportunities for experiments manipulating the 
quantum state of the system.

The coherence of the polariton condensate can be quantified by the
decay of the first-order coherence function, $\gf{\tau}\propto \langle
a^\dagger(0)a(\tau) \rangle$, whose Fourier transform is the emission
spectrum. For the polariton condensate, this function is directly
revealed by coherence measurements on the optical emission\
\cite{dima,kasprzak:067402,cond}.
In a condensate we expect long phase-coherence times and thus a spectrally
narrow emission above threshold. Recent experiments\ \cite{dima} have
shown that the decay time of \gf{\tau} is $\sim 150 \mathrm{ps}$, much
longer than was originally believed\ \cite{kasprzak:067402,cond}, but
short compared to a laser or atomic gas. Furthermore, the decay has a
distinctive Gaussian form. The experiments also determine the
intensity-intensity correlation function (second-order coherence
function), \gs{\tau}, which reveals significant number fluctuations
($\gs{0}>1$), decaying with a timescale $\sim 100 \mathrm{ps}$.

In this paper, we argue that the observed coherence properties of the
polariton condensate are produced by the combination of
polariton-polariton interactions, the number fluctuations generated by
the pumping, and a critical slowing-down of these number fluctuations
which occurs near the condensation threshold.  One or more of these
features are missing from previous theories of condensate coherence\
\cite{tassone00,thomsen,porras03,rubo03,laussy04,schwendimann08,eastham:085306},
yet all appear to be needed to obtain the behaviour now seen
experimentally. 
Rather than attempting a detailed microscopic model, our treatment is
based on general considerations of interacting, open
condensates. Since our predictions agree well with the experiment,
the model may be a useful guide in the development of 
microscopic theories\
\cite{tassone00,laussy04,porras03,rubo03,schwendimann08,doan08}.
The agreement with the experiments
was briefly described with the experimental results in Ref.[\onlinecite{dima}],
focusing on the case of slow number fluctuations appropriate for the
available data.  Here we provide a comprehensive treatment of the
theory, and show it predicts that other types of behaviour may also
be accessible in the experiments, most notably much longer coherence times
due to motional narrowing.

We first explain how the Kubo stochastic line-shape theory\
\cite{kubo,anderson} can be applied to give a general form for
\gf{\tau}, accounting for both interactions and number
fluctuations. This expression reduces to a Gaussian decay in the limit
where number fluctuations decay slowly. We then develop a solution to
a quantum-mechanical model of a pumped condensate, which gives the
same the Kubo form for \gf{\tau}. Close to threshold, this solution
gives a slow decay for number fluctuations, and hence explains the
observed Gaussian lineshape, as well as the observed decay
time. However, it also predicts that further above threshold number
fluctuations will be much faster, and significantly longer coherence times
should be obtained.

In our discussions, we neglect spatial effects, which is justified in the
CdTe system where the emission spot is strongly localised by disorder. We
thus model the condensate mode as a single anharmonic oscillator, 
with Hamiltonian
\begin{equation}
H= \adag a \, \omega_0 + \kappa \, (\adag a)^2,
\label{eq:ham}
\end{equation}
where $\omega_0$ is the oscillator frequency, and $\kappa$ the strength
of the polariton-polariton interactions. 

For a condensate of interacting particles, the interactions translate
number fluctuations in the condensate into random changes to its
energy, and so the coherence is lost. A similar effect, commonly
termed `self phase modulation', was originally observed for laser
beams propagating in a non-linear Kerr medium\ \cite{shimizu}. If we
assume that the condensate has a Gaussian probability distribution for
the number of polaritons with variance $\sigma^2$,
it is straightforward to obtain \gf{\tau} when the number fluctuations
are sufficiently slow\ \cite{dima}. It has a Gaussian form
\begin{equation}
|\gf{\tau}|=\exp{(-2 \kappa^2 \sigma^2 \tau^2)}
=\exp{(-\tau^2/\tau_c^2)}
\label{eq:gauss}
\end{equation}
As detailed in Ref.[\onlinecite{dima}], we can obtain, directly from
the measured data, values of $\sigma^2 \sim 25000$ and $\kappa \sim 2
\times 10^{-5}$ps$^{-1}$, giving a decay time $\tau_c \sim 200$ps, in
reasonable agreement with the experiments. The value for $\kappa$ is
consistent with theoretical estimates\
\cite{eastham:035319,marchetti:115326} of the interaction in a mode of
linear size $\sim 5 \mu \mathrm{m}$.

The picture described above is essentially static; it assumes that the
time-scale on which the number of polaritons changes, $\tau_r$, is much
longer than the coherence time $\tau_c$, so the only relevant time
evolution is caused by the action of the Hamiltonian. The obvious
problem with this description is that the coherence time, $\tau_c$, is
much longer than polariton life-time, $\tau_0 \sim 1{\rm ps}$, due to
emission from the cavity.  This suggests that the microcavity system
may well not be in the quasi-static regime, and we need to consider
the processes by which the number fluctuations occur in more detail.

The effect of introducing a time-scale, $\tau_r$, for fluctuations can
be understood using the Kubo stochastic line-shape theory\cite{kubo},
which describes the decay of \gf{\tau} for emission associated with a
transition whose energy varies randomly in time. In our case, the
random variations in energy are a consequence of the number
fluctuations, with time scale, $\tau_r$, determined by
the pumping. When the random energies have a Gaussian distribution,
the Kubo theory gives
\begin{equation}
|\gf{\tau}|=\exp{\left[-\frac{2 \tau_r^2}{\tau_c^2} 
(e^{-\tau/\tau_r}+\tau/\tau_r -1)
\right]}.
\label{eq:kubo}
\end{equation}
For $\tau_r$ slow compared to the scale set by the variance of the
random energy distribution ($\tau_c$ above), we are in the
static regime and a Gaussian decay with life time $\tau_c$ is
predicted. However, in the opposite regime, $\tau_r \ll \tau_c$,
motional narrowing occurs, and the decay becomes a simple exponential,
with life time $\tau_c^2/2 \tau_r$, much longer than $\tau_c$.  If we
naively take $\tau_{r}=\tau_0 \sim 2\mbox{ps}$, we would clearly be in
the motional narrowing regime, giving an exponential function, with a
very slow decay $\sim 10\mbox{ns}$. The measurement of the \gf{\tau}
decay thus shows that $\tau_r$ is, in fact, much longer than $\tau_0$,
and must be comparable to, or greater than, $\tau_c$. This prediction
is very well confirmed by the measurement of the decay of \gs{\tau},
which directly determines the time scale for number fluctuations; the
experimental decay time is $\sim 100\mbox{ps}$, similar to
$\tau_c$ for the same condensate population.

The explanation for the slow decay of \gs{\tau} comes from laser
physics, where it is well known that number fluctuations are slowed
close to the threshold. This can be explained using a simple classical
model where the pumping provides a gain which is saturable, that is,
has a dependence on the mode population.  Above threshold, the linear part of
the gain term exceeds the loss, so the population grows, and
a non-linear saturation term is required to obtain a finite steady
state. However, the response to small fluctuations, which is what the
intensity correlation experiment measures, depends only on the net
linear gain. Thus, near threshold, where the linear gain and loss are
closely matched, fluctuations in the system relax very
slowly. Haken\cite{haken} shows the close analogy of this behaviour to
the critical slowing down of fluctuations in the vicinity of an
equilibrium phase transition.

To put these considerations on a more formal footing, we have
developed a quantum model of the polariton condensate which can be
solved analytically for \gf{\tau} and \gs{\tau}. This model is a
generalisation of one studied by Thomsen and Wiseman\ \cite{thomsen},
in the context of atom lasers, to include the regime where motional
narrowing occurs; their model only applies to the `far above
threshold' limit, where the slowing of number fluctuations no longer
occurs. The coherent mode is treated as an anharmonic oscillator with
a Kerr non-linearity, Eq.\ (\ref{eq:ham}). This mode is coupled to a
reservoir, using the master equation formalism for the density matrix
$\rho$. Reservoir losses are offset by a standard laser-like saturable
pump term\ \cite{scully}. We thus obtain equations for the population
distribution, $P_n=\rho_{n,n}$, and the coherence $u_n=\sqrt{n}
\rho_{n-1,n} e^{-i
\omega_0 t}$:
\begin{equation}\label{eq:pop}\begin{split}
\dot{P_n} &= 
\gamma n_c \left[\frac{n}{n+n_s}P_{n-1}-\frac{(n+1)}{(n+1)+n_s}
P_n
\right] \\
&\quad + \gamma [(n+1) P_{n+1}-n P_n]
\end{split}
\end{equation}
and
\begin{equation}\begin{split}
\dot{u}_n &=
\gamma n_c \left[ 
\frac{n}{n+n_s-\mbox{\half}} u_{n-1}
-\frac{n+\mbox{\half}}{n+n_s+\mbox{\half}} u_n \right] \\ 
& \quad + \gamma \left[ n u_{n+1} - (n-\mbox{\half}) u_n \right] + 2 i
\kappa n u_n.\end{split}
\label{eq:pol}
\end{equation}
Here, $\half \gamma=1/\tau_0$ is the cavity decay rate, and $n_c$ and $n_s$
are parameters describing the pump process\footnote{In the terminology
of Ref.\ \cite{scully}, $\gamma={\cal C}$, $n_s={\cal A}/{\cal B}$,
and $n_c={\cal A}^2/{\cal BC}$. We have neglected some terms
$O(1/n_s)$, since $n_s \sim 10^4$ for our system.}. $n_c$
characterizes the strength of the pumping, while $n_s$ provides the
saturation: for $n \gtrsim n_s$ the gain decreases.  Physically, this
corresponds to the depletion of the pump reservoir by the processes
which populate the condensate.  Far above threshold, where the mean
occupation $\nav \gg n_s$, this model becomes identical to
Ref.~\cite{thomsen}. Pumping terms like this, at least to $O(n^2)$,
are required to give a finite condensate population, so the forms
(\ref{eq:pop}) and (\ref{eq:pol}), with quantitative $n_c$ and $n_s$,
should be derivable from microscopic treatments of the polariton
relaxation dynamics using kinetic equations~\cite{laussy04}.

The mean field threshold for condensation corresponds to $n_c=n_s$; for 
$n_c>n_s$, the steady state solution of Eq.(\ref{eq:pop}) is 
\begin{equation}
P_n^S \propto \frac{n_c^n}{(n+n_s)!}
\approx \exp{\left[-\frac{(n-\nav)^2}{2 n_c} \right]},
\label{eq:ss}
\end{equation}
where the Gaussian form is valid when $\nbar=n_c-n_s \gg
1$. The variance of the population is $\sigma^2=n_c$, so that number
fluctuations in the threshold region, $\nbar \ll n_s$, are
super-Poissonian. It is convenient to divide the threshold region into
two: for $\nbar \gtrsim 3\sqrt{n_s} \sim 500$, the Gaussian form is
valid for all $n$, because the non-physical $n<0$ states are not
significantly occupied, while for smaller values of \nbar, these
states have to be explicitly excluded. In the former case, the mean
population $\nav=\nbar$.

\begin{figure}
\begin{center}
\mbox{
\includegraphics[scale=0.35]{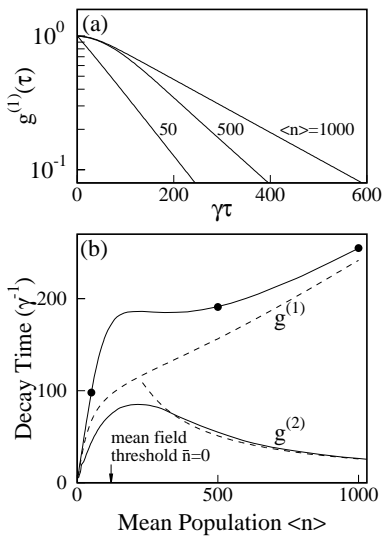}
}
\end{center}
\caption{
(a) The decay of \gf{\tau} for populations $\nav=$50, 500 and 1000.  On
this logarithmic plot, a simple exponential is linear, while a
Gaussian is quadratic.  (b) Decay times for \gf{\tau} and \gs{\tau},
as a function of population, obtained from the numerical solutions of
Eqs.(\ref{eq:pol}) and (\ref{eq:pop}) (solid lines). The nonlinearity
is $\kappa = 4 \times 10^{-5} \gamma$, and $n_s=2.5 \times 10^4$, as
derived from the experimental results. The dashed lines are from the
analytic expressions, Eqs.(\ref{eq:g1}) and (\ref{eq:gs}). The marked
points indicate the values of $\nav$ used in (a). }
\label{fig:decay}
\end{figure}

Eqs.(\ref{eq:pop}) and (\ref{eq:pol}) are easily solved numerically
for populations of a few hundred particles. Fig.(\ref{fig:decay}b)
shows the decay times for the correlation functions obtained from
these solutions, plotted as a function of the mean population. 
 As the population increases, the $g^{(2)}$ decay time
rises rapidly to $\sim 100\mbox{ps}$, then decreases. The $g^{(1)}$
time also rises quickly, then flattens for a while before increasing
again. Populations up to $\sim 500$ correspond to the experimental
regime of Ref.[\onlinecite{dima}], and the observed $g^{(1)}$
behaviour has a very similar form.  In Fig.(\ref{fig:decay}a), the
actual decay of \gf{\tau} is plotted, for three population values. For
$\nav=50$, corresponding to the initial rise in coherence time, the form
is a simple exponential. In the flat region, with $\nav=500$ the decay
starts off Gaussian, before becoming exponential at longer delays;
this is the near static behaviour of the self-phase modulation regime.
The final rise occurs when the $g^{(2)}$ time shortens and motional
narrowing sets in. This is evident in the decay curve for $\nav=1000$,
which starts off with the same Gaussian as $\nav=500$, but much sooner
becomes a slower exponential.

We now turn to deriving analytic solutions to Eqs.(\ref{eq:pop}) and
(\ref{eq:pol}), which provide the dashed lines on the figure. These
solutions are valid in the regime where $\nbar \gtrsim 3\sqrt{n_s}
\sim 500$ and the Gaussian in Eq.(\ref{eq:ss}) is valid without
truncating the $n<0$ part. The solutions include a term of
include a term of exactly the Kubo form Eq.(\ref{eq:kubo}), showing
that this is the correct description of the physics.

 Eq.\ (\ref{eq:pop}) for the
population is independent of the non-linearity, so we can quote 
standard laser theory results~\cite{scully} for the intensity
correlation function:
\begin{equation}
\gs{\tau}=1+\frac{n_s}{\nbar^2} 
\exp{\left(-\frac{\nbar}{n_c} \gamma \tau \right)}
=
1+\frac{n_s}{\nbar^2} \exp{(-\gbar \tau)},
\label{eq:gs}
\end{equation}
where $\gbar=\nbar \gamma / n_c$ is the slowed decay rate. This 
result fits the single experimental data point fairly well; using the
experimental measurements $\gs{0}=1.1$ and $\nbar=500$ we obtain
$n_s=25000$ and $\gbar=\gamma/50$. This gives a decay time of $\sim
50 \mbox{ps}$, in reasonable agreement with the measured 100ps.

The first order correlation function, $\gf{\tau} \propto 
\sum_n u_n(\tau) e^{i \omega_0 \tau}$, is
obtained by solving Eq.\ (\ref{eq:pol}). The solution is required with
an initial condition $u_n(0)=nP_n^{S}$, which is a similar Gaussian
function to $P_n^S$, but with mean $\nbar+n_c/\nbar$.  To obtain the
correlation function, we follow the approach of Gardiner and Zoller~
\cite{gardiner}. 
Using a Kramers-Moyal expansion, the difference
operators are converted into differentials, leading to a
Fokker-Planck equation for $u$, which we now write as $u(n,t)$, with
$n$ a continuous variable. To deal with the appearance of $n$ in the
denominator of the pumping terms, we linearise around the mean value,
writing $n=\nbar+n_c/\nbar+n'$. Thus we obtain
\begin{multline}
\frac{\partial u}{\partial t}
=
2i \left( \kappa + \frac{i \gamma}{4 n_c} \right) n' \, u
- \frac{\gamma}{2 \nbar} u \\  + 
\left( \gamma \frac{\nbar}{n_c} \right) 
\frac{\partial}{\partial n'} 
\left[
n' \, u
+\half
(2 n_c)
\frac{\partial u }{\partial n'}
\right], \end{multline}
where we have omitted constant nonlinear contributions to the oscillator 
energy, which can be absorbed in a renormalised $\omega_0$. This equation
is solved in the Fourier domain\cite{gardiner} to give
\begin{multline}
|\gf{\tau}|=
\exp{\left[-
\frac{4 n_c \kappa^2}{\gbar^2} {(e^{-\gbar \tau}+\gbar \tau -1)}
\right]} \\ \times
\exp{\left[
\frac{n_c}{4 \nbar^2} {(e^{-\gbar \tau}- \gbar \tau -1)}
\right]}.
\label{eq:g1}
\end{multline}
The first factor is just the Kubo expression, Eq.\ (\ref{eq:kubo}),
with $\tau_c^2=1/2 \kappa^2 \sigma^2$ and $\sigma^2=n_c$, as before,
and $\tau_r=1/\gbar$. This constitutes the main result of our
treatment: we obtain Kubo type behaviour, with the fluctuation time
$\tau_r$ given by the decay time of \gs{\tau}. The second term
corresponds to the Schawlow-Townes decay, enhanced in the threshold
region due to the finite amplitude fluctuations. It is generally much
slower than the first, in the regime where the expression is valid.

The figure shows very clearly the importance of the fluctuation time
scale $\tau_r \propto 1/\nbar$ (for large \nbar) on the coherence
time. When \nbar\/ is increased, $\tau_c$ only changes very slightly,
in fact decreasing as $n_c=n_s+\nbar$ grows, but $\tau_r$ shortens
rapidly. This pushes the system into the motional narrowing regime,
where the life time $\tau_c^2/2 \tau_r$ is proportional to \nbar. Note
that we have to be careful treating this as a prediction of a linear
relationship between coherence time and emission intensity for high
powers; we have simply kept $n_s$ constant and increased $n_c$.  In
the textbook laser model\ \cite{scully}, $n_s$ is indeed independent
of pump power $P$, and $n_c \propto P$. Though pumping of the the
polariton system is considerably more complex than this, it is
likely that $n_c$ increases more rapidly with $P$ than $n_s$
does. Thus the fluctuation decay should decrease with
stronger pumping, potentially taking the condensate into the
motional narrowing regime. For this to be observable, it would, of
course, be necessary that other mechanisms should not take over and
restrict the coherence as the number fluctuation effect is suppressed.

One surprising feature of the experiments is that very significant
slowing down is still occurring at pump powers of twice the threshold
value, $P_{th}$. This is inconsistent with the simplest assumption,
$n_c \propto P$, since then $\bar \gamma/\gamma = (1-P_{th}/P)$ is
close to one at these powers. A full understanding of this requires
microscopic models of the pumping, but it may be explained by assuming
that there are a limited number of reservoir states which can provide
gain for the condensate, so the gain parameter $n_c$ must become
independent of $P$ at high pumping. In a system where the
maximum achievable gain only just exceeds the loss, the reservoir must
be almost full at threshold, so this saturation will extend the
critical region to large values of $P/P_{th}$.

Instead of increasing the pump power, an alternative way of obtaining
larger populations and hence longer coherence times may be to increase
the size of the condensate. In the experiments, the size is
determined by the disorder, so this means finding larger emission
spots.  We present a simple argument for how the various time scales
should vary with spot size, $A$, assuming constant density, that is
$\nbar \propto A$. Since $\nbar=n_c-n_s$, this suggests $n_c \sim A^1$
and $n_s \lesssim A^1$, for large $A$. The polariton lifetime,
$\tau_0$ is independent of $A$, while $\kappa \sim A^{-1}$.  Together,
this gives $\tau_c \sim \sqrt{A}$ and $\tau_r$ independent of
$A$. Hence, in the static regime, the coherence time, which is then
just $\tau_c$, would grow as $\sqrt{A}$, but for sufficiently large
$A$, this will always take us in the motional narrowing regime,
$\tau_c \gg \tau_r$, and the coherence time will be $\tau_c^2/\tau_r
\sim A$. This argument shows that the decoherence is a mesoscopic
phenomenon; as in other ordered phases such as ferromagnets and
lasers, the finite condensate has a finite ergodic time, beyond which
the order-parameter correlations disappear. However, in the
thermodynamic limit $A \rightarrow \infty$, this ergodic time
diverges, and true symmetry breaking can occur.

In summary, we have shown that the available experimental results are
well explained by a model of a microcavity polariton condensate as a
pumped, dissipative system where the main decoherence process is the
combined effect of number fluctuations and inter-particle
interactions. However, our model also predicts that a regime of
motional narrowing should be accessible at higher pump powers, which
would lead to much longer coherence times. This would make the
polariton condensate much more suitable for experiments involving the
manipulation of its quantum state, such as generating Josephson oscillations,
ultimately leading to applications in quantum information processing.

\begin{acknowledgments}
We gratefully acknowledge numerous discussions with Dmitry
Krizhanovskii, Maurice Skolnick and Pieter Kok, which were very
valuable in the development of this work. PRE acknowledges support
from EPSRC-GB Grant No. EP/ C546814/01.
\end{acknowledgments}


\end{document}